\documentclass{jfm}

\usepackage{flushend}
\usepackage{graphicx}
\usepackage{amsmath}
\usepackage{pgfplots}

\usepackage[normalem]{ulem}

\usepackage{siunitx}

\shorttitle{Settling and Clustering of Snow Particles in Atmospheric Turbulence}
\shortauthor{C. Li and others}

\title{Settling and Clustering of Snow Particles in Atmospheric Turbulence}

\author{Cheng Li\aff{1,}\aff{2}, Kaeul Lim\aff{1,}\aff{2}, Tim Berk\aff{1,}\aff{3}, Aliza Abraham\aff{1,}\aff{2}, Michael Heisel\aff{1,}\aff{4}, Michele Guala\aff{1,}\aff{4}, Filippo Coletti\aff{1,}\aff{3}, Jiarong Hong\aff{1,}\aff{2}\corresp{\email{jhong@umn.edu}}
		}
	
	\affiliation{\aff{1}St. Anthony Falls Laboratory, University of Minnesota, Minneapolis, MN 55414, USA
		\aff{2}Department of Mechanical Engineering, University of Minnesota, Minneapolis, MN 55455, USA
		\aff{3}Department of Aerospace Engineering and Mechanics, University of Minnesota, Minneapolis, MN 55455, USA
		\aff{4}Department of Civil, Environmental, and Geo- Engineering, University of Minnesota, Minneapolis, MN 55455, USA
	}

\begin{document}

\maketitle   

\begin{abstract}	
The effect of turbulence on snow precipitation is not incorporated into present weather forecasting models. Here we show evidence that turbulence is in fact a key influence on both fall speed and spatial distribution of settling snow. We consider three snow fall events under vastly different levels of atmospheric turbulence. We characterize the size and morphology of the snow particles, and we simultaneously image their velocity, acceleration, and relative concentration over vertical planes about 30~m$^2$ in area. We find that turbulence-driven settling enhancement explains otherwise contradictory trends between the particle size and velocity. The estimates of the Stokes number and the correlation between vertical velocity and local concentration indicate that the enhanced settling is rooted in the preferential sweeping mechanism. When the snow vertical velocity is large compared to the characteristic turbulence velocity, the crossing trajectories effect results in strong accelerations. When the conditions of preferential sweeping are met, the concentration field is highly non-uniform and clustering appears over a wide range of scales. These clusters, identified for the first time in a naturally occurring flow, display the signature features seen in canonical settings: power-law size distribution, fractal-like shape, vertical elongation, and large fall speed that increases with the cluster size. These findings demonstrate that the fundamental phenomenology of particle-laden turbulence can be leveraged towards a better predictive understanding of snow precipitation and ground snow accumulation. They also demonstrate how environmental flows can be used to investigate dispersed multiphase flows at Reynolds numbers not accessible in laboratory experiments or numerical simulations.
\end{abstract}


\section{Introduction}
The fall speed of snow, and frozen hydrometeors in general, is a crucial parameter for meteorological prediction \citep{Hong2004}.
The spatio-temporal distribution of snow precipitation directly impacts the ground accumulation, which in turn influences the local hydrology, road conditions, vegetation development, avalanche danger, and mass balance of glaciers \citep{Lehning2008,Scipion2013}.
At a global level, the velocity at which ice and snow particles settle in the atmosphere is one of the most important determinant of climate sensitivity \cite[][Fifth assessment report]{IPCC2014}.
Considering its importance, our understanding of snow particle settling is far from satisfactory, and the process remains poorly characterized \citep{Heymsfield2010a}.
We will use the word settling as it is more common in fluid mechanics, although the atmospheric science literature often terms it sedimentation.
Also, because we focus here on relatively small hydrometeors as opposed to large dendritic ones, we will generally refer to snow particles as opposed to snowflakes.

A common approach is to parameterize the vertical velocity of the snow particles (or generic hydrometeor) $W_s$ as a power-law function of the characteristic diameter $d_s$ \citep{Locatelli1974}:
\begin{equation}
W_s = a_w d_s^{b_w} \label{eq:eq1}
\end{equation}
\noindent where $a_w$ and $b_w$ are empirical constants.
A first difficulty lies in the specificity of the constants to the type of hydrometeors, which display a broad range of size, morphology, porosity, and riming that affect the balance between drag and gravity \citep{Pruppacher1997}.
Fall speed relationships that include the object mass $m_s$ and frontal area $A_s$ enable the definition of a particle Reynolds number and drag coefficient, and show more generality \citep{Boehm1989,Mitchell1996}.
However, even these models are ultimately similar to equation~\ref{eq:eq1} in that they resort to a power-law dependence of $m_s$ and $A_s$ on $d_s$.

Small snow particles and ice crystals (mm-sized or smaller) form the vast majority of the frozen precipitation in the atmosphere \citep{Pruppacher1997}.
Field studies focused on these small particles report values of the coefficient $b_w \approx 0.25$ (although with significant variability, see \citealt{Locatelli1974,Tiira2016,VonLerber2017}).
From the fluid mechanics standpoint, this would seem a very weak relation between the fall speed and the diameter.
Stokes drag implies $b_w = 2$, and while non-linear drag certainly affects the process, it is not expected to be dominant as the particle Reynolds number for those small hydrometeors is typically $\mathcal{O}(10)$.
Important factors that contribute to the trend include: particle bulk density, which varies significantly with the level of riming and porosity \citep{Pruppacher1997}; particle anisotropy, which can be high for needles and crystal aggregates \citep{Dunnavan2019}; and in general the complexity of the snow particle morphology, especially for dendritic ice crystals, aggregates, and open geometries \citep{Westbrook2008,Heymsfield2010a}.
Morphological factors, however, are not expected to play major roles for small hydrometeors of compact shape, while the weak dependence with the diameter is still observed \citep{Tiira2016}.
Therefore, it is evident that other environmental factors can influence the settling process besides the snow particle properties.

The role of atmospheric turbulence on the snow fall speed has only recently been recognized.
\citet{Garrett2014} considered data from a field study and showed that, in high turbulence, the fall speed was seemingly insensitive to the snow particle diameter.
They simultaneously measured hydrometeor morphology and fall speed using a multi-camera system, and found that unrealistic density estimates were obtained by assuming that the observed vertical velocity coincided with the terminal velocity in quiescent air.
While it is intuitive that air turbulence would broaden the distribution of snow particle velocities (as recently confirmed, \citealt{Garrett2015}), one may also expect the effectively Gaussian velocity fluctuations to cancel out, leaving the average fall speed unaffected. This view, however, does not account for well-known phenomena in particle-laden flows.

When small heavy particles fall through turbulence, their mean settling velocity can be significantly altered compared to still-fluid conditions \citep{Nielsen1993,Wang1993,Balachandar2010}.
The fall can be hindered, e.g., if weakly inertial particles are trapped in vortices \citep{Tooby1977}, or if fast-falling particles are slowed down by non-linear drag \citep{Mei1991} or loiter in upward regions of the flow \citep{Good2012}.
More often, however, turbulence is found to enhance the settling through a process known as preferential sweeping \citep{Maxey1987,Wang1993}: inertial particles favour downward regions of the flow, i.e., they oversample fluid with vertical velocity fluctuations aligned with the direction of gravity.
This effect is especially strong (and dominant over other mechanisms that hinder the fall) when the particles have an aerodynamic response time $\tau_p$ comparable to the Kolmogorov time scale $\tau_\eta$; that is, when the Stokes number $St \equiv \tau_p/\tau_\eta = \mathcal{O}(1)$.
Laboratory experiments have shown that in this case the mean vertical velocity can see a multi-fold increase \citep{Aliseda2002,Good2014,Huck2018,Petersen2019}.
Recently, \citet{Nemes2017} imaged and tracked snow particles in the atmospheric surface layer, observed a substantial increase in settling velocity, and suggested that preferential sweeping was at play.

Another well-known behaviour exhibited by inertial particles in turbulence is the tendency to form clusters, especially when $St = \mathcal{O}(1)$ \citep{Eaton1994,Monchaux2012,Gustavsson2016}.
Along with the ability of inertial particles to maintain significant relative velocity for vanishing separations, this effect is thought to enhance their collision rate \citep{Sundaram1997,Bewley2013}.
As such, clustering is expected to be consequential for a variety of natural phenomena, from atmospheric cloud dynamics \citep{Shaw2003,Grabowski2013} to dust agglomeration in circumstellar nebulas \citep{Cuzzi2001}.
Precipitating snow can be largely composed of aggregates formed by successive collisions of ice crystals \citep{Dunnavan2019}.
Thus, if clustering of frozen hydrometeors does occur, it is likely to play an important role in determining their particle shape, size and fall speed.
To date, there is no direct evidence that snow particles cluster in the atmosphere, nor the properties that such clusters may possess, and the impact turbulence may have on the evolution of frozen precipitation remains speculative.

Here we present and analyse data from three field studies where settling snow is illuminated and imaged over vertical planes ${\sim} \: 30$~m$^2$, using previously introduced approaches \citep{Hong2014,Nemes2017}.
We characterize the snow particle velocities and accelerations across a broad range of atmospheric conditions, and show that turbulence plays a dominant role in determining both mean and variance of the snow fall speed.
Moreover, we document for the first time the appearance of clusters in the snow spatial distribution, describe their multi-scale geometry and assess their settling velocity.
The paper is organized as follows: the experimental setups to characterize the atmospheric 
conditions, snow particle properties, and large-scale velocity fields are described in \S\ref{sec:sec2}; the results in terms of snow particle size, velocity, acceleration, and concentration fields are reported in \S\ref{sec:sec3}; in \S\ref{sec:sec4} we draw conclusions and discuss future perspectives.

\section{Methodology} \label{sec:sec2}
\subsection{Field experiment setups}
The data presented in the current study were acquired in three field deployments conducted at the Eolos Wind Energy Research Field Station in Rosemount, MN, between 2016 and 2019.
We will refer to them as Jan2016, Nov2018, and Jan2019.
The station features a meteorological tower instrumented with wind velocity, temperature, and humidity sensors installed at elevations ranging from 7~m to 129~m.
Four of these elevations (10, 30, 80 and 129~m) are instrumented with Campbell Scientific CSAT3 3D sonic anemometers with a sampling rate of 20~Hz.
Detailed descriptions of the field station and instrument specifications are provided in \citet{Hong2014} and \citet{Toloui2014}. 

For each deployment, the spatial distribution and motion of the settling snow is captured using super-large-scale particle image velocimetry (PIV) and particle tracking velocimetry (PTV) described in \citet{Hong2014} and \citet{Nemes2017}, respectively.
The size and shape of snow particles are simultaneously obtained using an in-house digital in-line holographic (DIH) system introduced in \citet{Nemes2017}.
A minimum of 15000 holograms is captured for each dataset.
Because the DIH setup is located just meters away from the PIV/PTV field of view, spatial variability between both measurements is deemed negligible.
The key information for the PIV/PTV and DIH systems in each deployment is summarized in table~\ref{tab:tab1}, and further details on the setups are given in the following.
All three sets have fairly constant snowfall and wind intensity.

\begin{figure}
	\centering
	\includegraphics{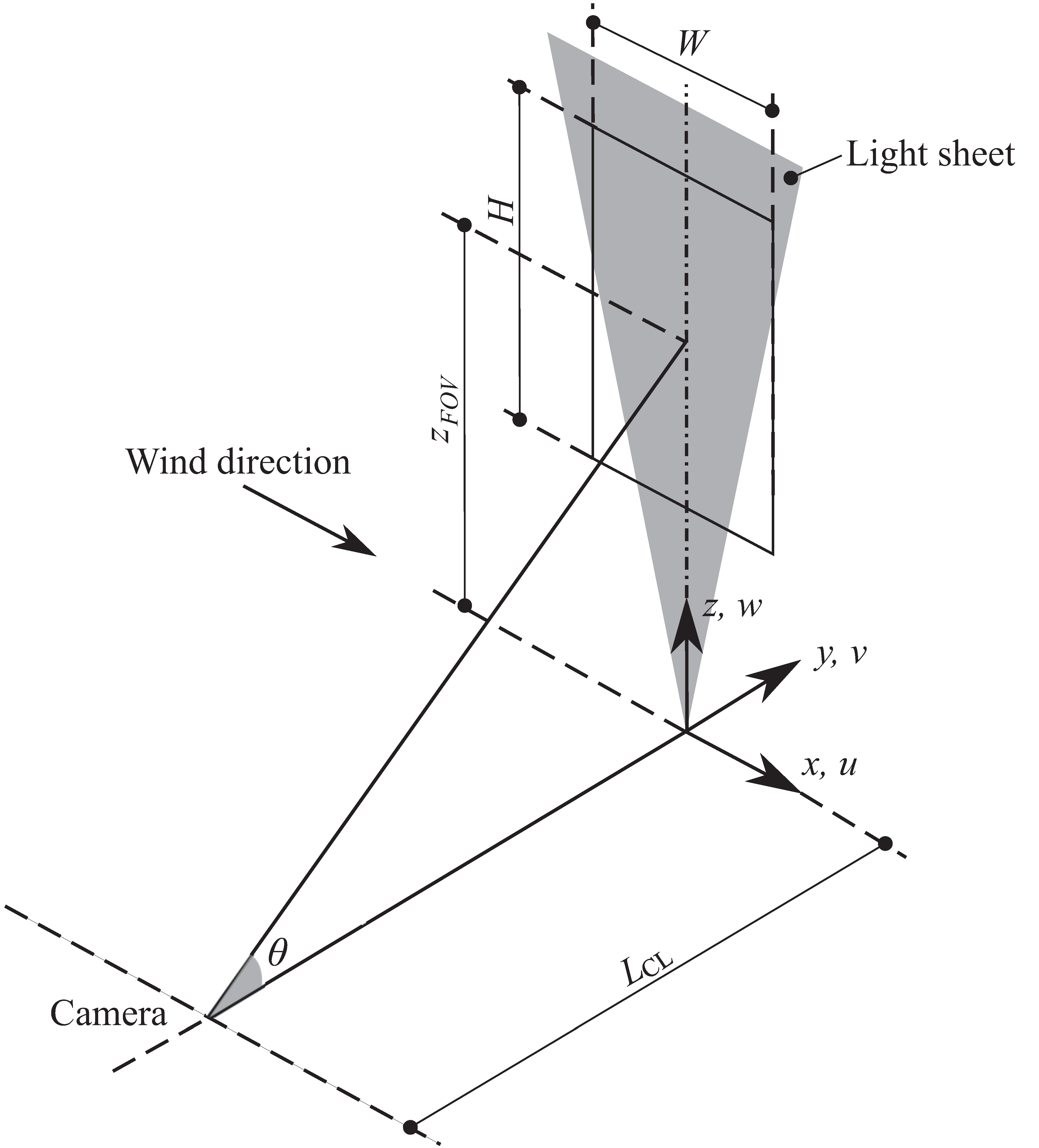}
	\caption{Schematic of the measurement setup used in the deployments. The field of view (FOV) has width $W$, height $H$, and is centered at an elevation $z_{FOV}$ (see table~\ref{tab:tab1}). The other symbols are defined in the text.
	}
	\label{fig:fig1}
\end{figure}

The PIV/PTV setup is similar to the one used in \citet{Nemes2017}.
The illumination is provided by a 5-kW search light with a divergence $<0.3$ degrees and an initial beam diameter of 300~mm, shining on a curved mirror that redirects the beam vertically and expands it into a light sheet.
The system is attached to a trailer for mobility in aligning the sheet with the wind direction and minimizing out-of-plane motion.
The camera is placed on a tripod at a distance $L_{CL}$ from the light sheet with a tilt angle $\theta$ from the horizontal.
The coordinate system (streamwise $x$, spanwise $y$, and vertical $z$, and the corresponding velocity components $u$, $v$, $w$) as well as the position and dimensions of the field of view (FOV) are illustrated in figure~\ref{fig:fig1}.

\begin{table}
	\centering
	\begin{tabular}{c c c c c c c c c}
		& \multicolumn{6}{c}{PIV/PTV Setup} & \multicolumn{2}{c}{DIH Setup}\\ \cline{2-7}
		& & & & & & & & \\
Dataset & Duration & $z_{FOV}$ & $H \times W$ & Resolution & $\theta$ & $L_{CL}$ & Resolution & Volume \\
& (minutes) & (m) & (m$^2$) & (mm/pixel) & (degrees) & (m) & (\si{\micro\meter}/pixel) & (cm$^3$)\\
		& & & & & & & & \\
		Jan2016 & 5 & 10.8 & $7.1 \times 4.0$ & 5.6 & 21.1 & 25 & 24 & 18.8\\
		Nov2018 & 17 & 9.1 & $8.4 \times 4.7$ & 6.5 & 14.5 & 31 & 14 & 42\\
		Jan2019 & 15 & 20.2 & $14.7 \times 8.3$ & 12.0 & 19.9 & 53 & 14 & 42\\
	\end{tabular}
	\caption{Summary of key parameters of particle image velocimetry (PIV), particle tracking velocimetry (PTV) and digital inline holography (DIH) measurement setups for each deployment dataset used in the present paper (see figure~\ref{fig:fig1}). All PIV/PTV datasets have the same acquisition rate of 120~fps.}
	\label{tab:tab1}
\end{table}

\begin{figure}
	\centering
	\includegraphics{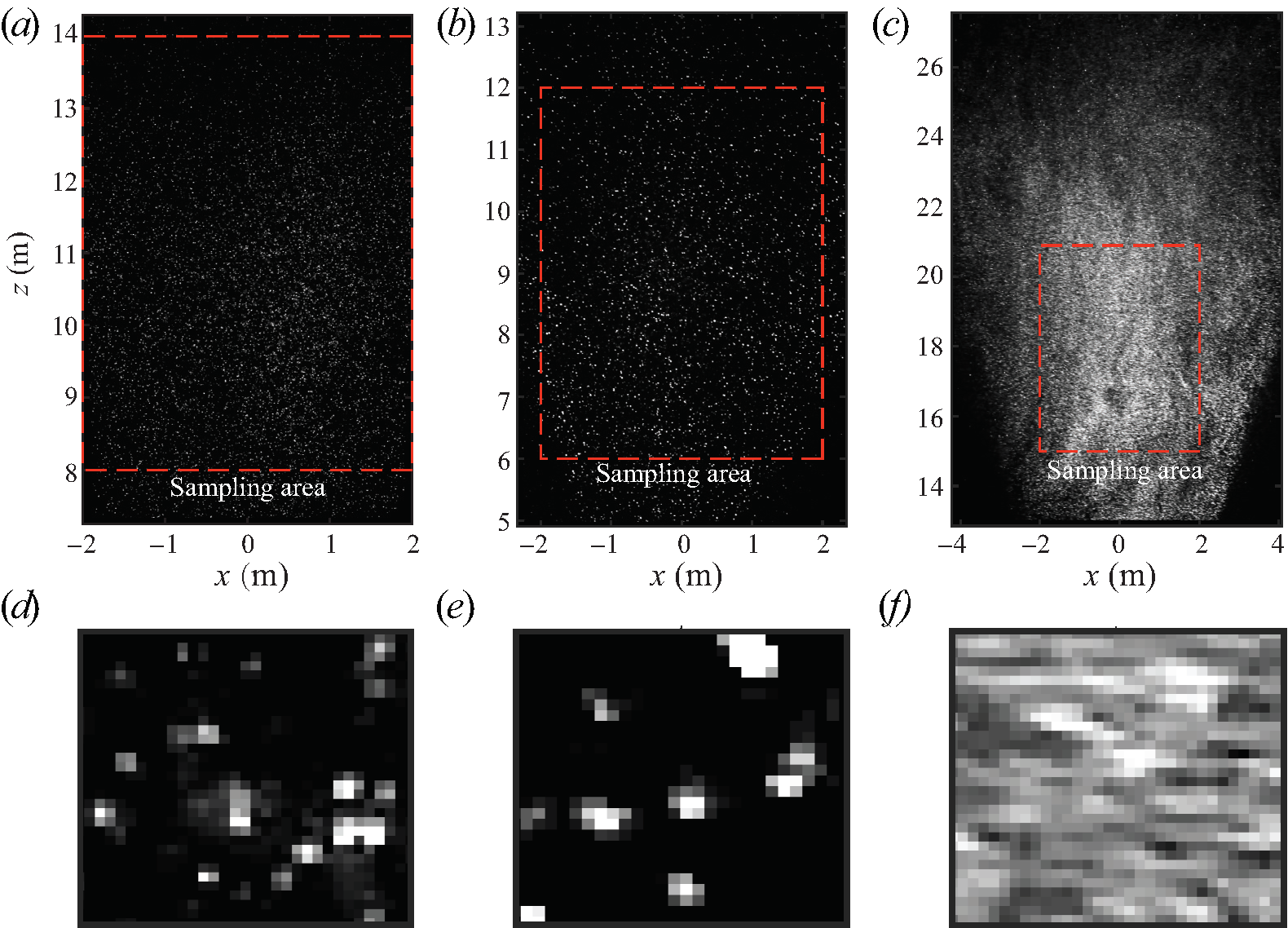}
	\caption{Samples of raw snow particle images at the full field of view (a-c) used for PIV/PTV, and close-up on the 32 $\times$ 32 pixels PIV interrogation window (d-f) from datasets Jan2016 (a, d), Nov2018 (b, e), and Jan2019 (c, f).
	}
	\label{fig:fig2}
\end{figure}

Figure~\ref{fig:fig2} shows sample images used for PIV/PTV in each deployment, which span a broad range of imaging conditions and have different FOVs (figure~\ref{fig:fig2}a--\ref{fig:fig2}c).
For consistency, a sampling region of 7~m $\times$  4~m (matching the Jan2016 FOV) with the same 32~pixel $\times$ 32~pixel PIV interrogation window (figure~\ref{fig:fig2}d--\ref{fig:fig2}f) is used to analyse all the datasets.
Based on previous parameterisation of the boundary layer at the Eolos site \citep{Heisel2018}, the selected regions are located in the logarithmic layer, well above the roughness sublayer and possible snow saltation layer \citep{Guala2008}.
The particle image density is similar for Jan2016 and Nov2018, and significantly higher for Jan2019.
Accordingly, both PIV and PTV are applied to Jan2016 and Nov2018, while only PIV (which can deal with high particle image densities) is performed on Jan2019 dataset.
PIV provides Eulerian velocity fields (for atmospheric measurements, see \citealt{Hong2014,Toloui2014,Heisel2018}), which we use to measure the snow fall speed, while PTV provides Lagrangian trajectories, which we use to calculate snow particle accelerations \citep{Nemes2017}.
Detailed information on the image and data processing (e.g., distortion correction, background subtraction, PIV cross-correlation, Lagrangian particle tracking) was previously reported \citep{Hong2014,Toloui2014,Nemes2017,Dasari2018}.
The same images are also used to estimate the relative snow particle concentration, as described in \S\ref{sec:sec3.3}.

Two versions of the DIH system are employed to characterize the snow particle size, shape, and number density.
The earlier version is employed for Jan2016 and is described in detail in \citet{Nemes2017}.
The later version, which has larger sampling volume and improved spatial resolution and data acquisition capabilities, is used for Nov2018 and Jan2019.
It uses a diode laser (Roithner 5~mW, wavelength of 635~nm), a beam expander (Edmund Optics 9~mm plano-concave lens), and a collimating lens (Thorlabs 100~mm biconvex lens with anti-reflective coating) to generate a 50~mm beam. A CMOS camera (PtGrey Blackfly 2048 $\times$ 1536~pixels, 3.45~\si{\um}/pixel), mounting a Fujinon 25~mm f/1.4 lens, captures the holograms resulting from the interference of the light scattered by the snow particles and the collimated beam.
The camera is connected to a data acquisition system (Raspberry Pi 3 - Model B), which interfaces with a laptop running FLIR SpinView software to control the camera and collect the images.
Both versions of the DIH system are mounted about 2~m above ground level and allow the snow particles to settle through the sampling volume with minimal disturbance.
\citet{Nemes2017} describe the processing steps through which detailed two-dimensional projections of the snow particle silhouette are obtained from the holograms.

\subsection{Meteorological conditions and turbulence properties}
\begin{figure}
	\centering
	\includegraphics{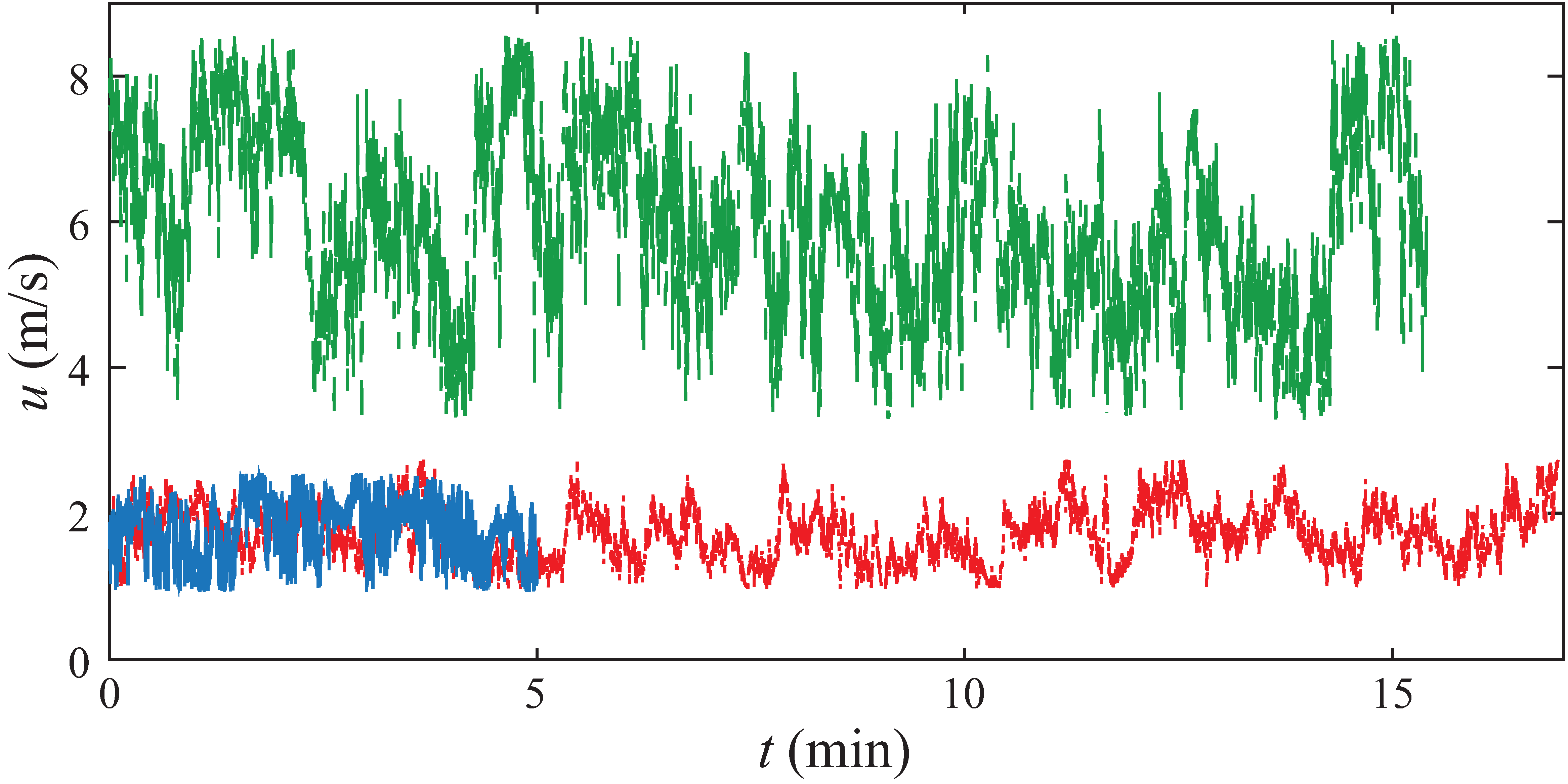}
	\caption{Time series of the sonic anemometer data at elevation $z=10$~m showing streamwise wind velocity for Jan2016 (blue), Nov2018 (red), and Jan2019 (green).
	}
	\label{fig:fig3}
\end{figure}

Simultaneous measurements from the meteorological tower sensors provide a statistical description of the turbulence conditions during the PIV/PTV and DIH measurements.
As shown in figure~\ref{fig:fig3}, the time series of the wind velocity ($u$) and fluctuation ($u'$) from the 10~m sonic anemometer (at an elevation comparable to the PIV/PTV FOV) indicate good stationarity for all datasets.
Table~\ref{tab:tab2} summarizes the key meteorological and turbulence parameters.
Atmospheric stability conditions are estimated based on both the bulk Richardson number $R_b$ and the Monin-Obukhov length $L_{OB}$:
\begin{equation}
R_b = -|g|\Delta\overline{\theta_v} \Delta z / \left(\overline{\theta_v}\left[(\Delta U)^2 + (\Delta V)^2\right]\right)
\end{equation}
\begin{equation}
L_{OB} = -U_\tau^3 \overline{\theta_v}/\kappa g \overline{w'\theta_v'}
\end{equation}

\begin{table}
\centering
\begin{tabular}{c c c c c c c c c c c c c c}
	Dataset & $U$ & $u_{rms}$ & $RH$ & $T$ & $R_b$ & $L_{OB}$ & $L$ & $\tau_L$ & $\varepsilon$ & $\eta$ & $\tau_\eta$ & $Re_\lambda$ & $G\tau_\eta$ \\
	& (m/s) & (m/s) & (\%) & ($^\circ C$) & (-) & (m) & (m) & (s) & (cm$^2$s$^{-3}$) & (mm) & (ms) & (-) & (-) \\
	Jan2016 & 1.98 & 0.16 & 98.3 & -0.1 & 0.03 & 811 & 4.9 & 30.6 & 8 & 1.29 & 126 & 938 & 0.003\\
	Nov2018 & 1.55 & 0.38 & 94.3 & -1.8 & 0.12 & 1007 & 3.4 & 8.9 & 19 & 1.04 & 83 & 3545 & 0.003\\
	Jan2019 & 5.95 & 1.18 & 80.0 & -16.0 & 0.03 & -2643 & 14.6 & 12.9 & 290 & 0.49 & 20 & 9180 & 0.008\\
\end{tabular}
\caption{The meteorological and turbulence parameters obtained using the sonic anemometer at $z=10$~m for all three datasets. See the text for the definition of the symbols.}
\label{tab:tab2}
\end{table}

Here $g$ is the gravitational acceleration, $\theta_v$ is the virtual potential temperature (calculated using 1-Hz temperature, pressure, and relative humidity measurements), $U$ and $V$ are the average North and West wind velocity components, respectively, $U_\tau$ is the shear velocity, $\kappa$ is the von K\'{a}rm\'{a}n constant, prime indicates temporal fluctuations and the overbar indicates time-averaging.
The mean velocity differences $\Delta U$ and $\Delta V$ are measured from the sonic anemometers at 10~m and 129~m ($\Delta z=119$~m).
The length $L_{OB}$  and all other turbulence quantities reported in this section are evaluated from the 20~Hz sonic sensor at $z=10$~m.
We approximate the surface turbulent virtual potential heat flux with the measured $\overline{w'\theta_v'}$.
The friction velocity $U_\tau$ is estimated based on the Reynolds stresses \citep{Stull1988}:
\begin{equation}
U_\tau = \left(\overline{u'w'}^2 + \overline{v'w'}^2\right)^{1/4}
\end{equation}

For all deployments, $R_b \ll 1$ and $z/L_{OB} \ll 0.1$, indicating that the boundary layer flow within the measuring domain can be approximated as neutrally stratified \citep{Hoegstroem2002}.

The integral time scale $\tau_L$ and length scale $L$ are calculated from the temporal auto-correlation function $\rho$:
\begin{equation}
\rho(\tau) = \overline{u'(t)u'(t+\tau)}/\overline{u'(t)^2}
\end{equation}
\begin{equation}
\tau_L = \int_{0}^{T_0} \rho(\tau) \mathrm{d}\tau
\end{equation}
\begin{equation}
L = u_{rms} \tau_L
\end{equation}
where $t$ is time, $\tau$ is the temporal separation, and $T_0$ is the first zero-crossing point of the auto-correlation function.
Here and in the following, rms indicates root mean square fluctuation.
The turbulent dissipation rate $\varepsilon$ for Nov2018 and Jan2019 is estimated using the second-order temporal structure function of the streamwise velocity fluctuations:
\begin{equation}
D_{11}(\tau) \equiv \overline{\left[u'(t+\tau)-u'(t)\right]^2}
\end{equation}

\begin{figure}
	\centering
	\includegraphics{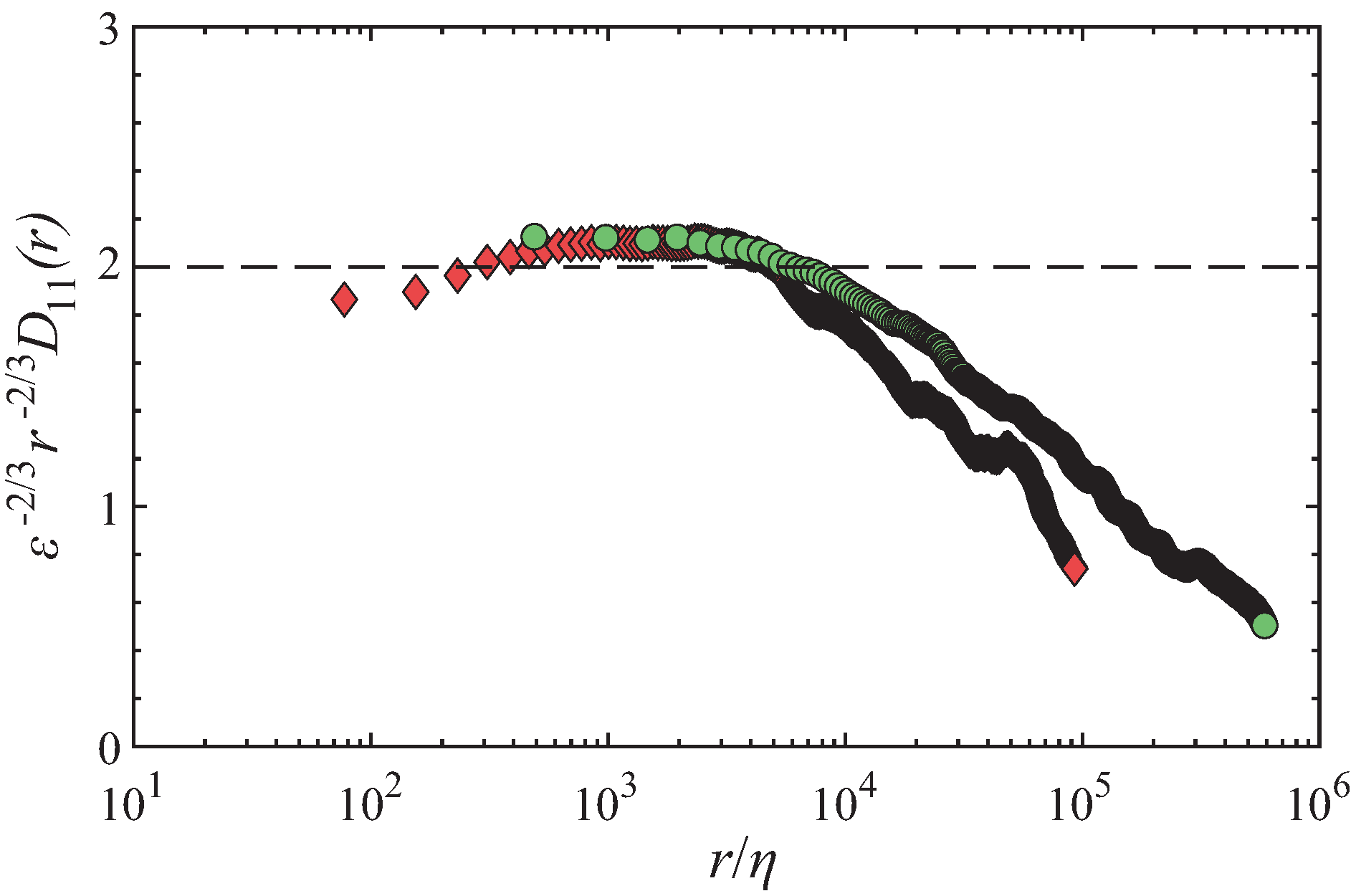}
	\caption{Compensated second-order structure function of the streamwise velocity fluctuations calculated from the sonic anemometer for Nov2018 (red diamonds), and Jan2019 (green circles) at $z=10$~m. The dashed line indicates the inertial range prediction of equation~\ref{eq:eq2.8} with $C_2 = 2$.
	}
	\label{fig:fig4}
\end{figure}

To yield better convergence of $D_{11} (\tau)$, the velocity time series are divided into two-minute windows with 50\% overlap.
Invoking Taylor hypothesis, the temporal separation is converted to a spatial separation $r=\tau U_{2\mathrm{min}}$, where $U_{2\mathrm{min}}$ is the mean velocity in each two-minute window.
We then calculate $\varepsilon$ using the Kolmogorov prediction for the spatial second-order structure function in the inertial range:
\begin{equation}
D_{11}(r) = C_2(\varepsilon r)^{2/3} \label{eq:eq2.8}
\end{equation}
where C$_2$ is a constant close to 2 in high-Reynolds number turbulence \citep{Saddoughi1994}.
The compensated structure functions in figure~\ref{fig:fig4} show good agreement with equation~\ref{eq:eq2.8} for a broad range of separation time scales in both Nov2018 and Jan2019 datasets.
For Jan2016 the convergence of the structure function is less satisfactory and the dissipation is approximated from classic scaling arguments, i.e., $\varepsilon = u_{rms}^3/L$.
We then obtain the Kolmogorov time and length scale, $\tau_\eta = \left(\nu/\varepsilon\right)^{1/2}$ and $\eta = (\nu^3/\varepsilon)^{1/4}$, respectively, where $\nu$ is the air kinematic viscosity.
The Reynolds number $Re_\lambda = u_{rms}\lambda/\nu$ (where $\lambda = u_{rms}(15\nu/\varepsilon)^{1/2}$ is the Taylor microscale) spans a full decade across the three deployments, allowing us to investigate turbulence of vastly different intensity.
The mean shear across the field of view $G=\partial U/\partial z$ is much smaller than the small scale velocity gradients, i.e. $G(\nu/\varepsilon)^{1/2} = G\tau_\eta \ll 1$, and we thus expect approximate small-scale isotropy \citep{Saddoughi1994}.
This enables the comparison with previous laboratory experiments and simulations of particle-turbulence interactions performed in (nearly) homogeneous isotropic turbulence, consistent with the approach of \citet{Nemes2017}.

\section{Results} \label{sec:sec3}
\subsection{Snow particle size and settling velocity}
Table~\ref{tab:tab3} provides a summary of the average size, aspect ratio, and concentration of the snow particles as measured by DIH.
The particle size $d_s$ is quantified using the projected-area diameter, corresponding to the diameter of the circle with the same projected area as the particle image.
The aspect ratio $s_2/s_1$ is defined as the ratio between minor and major axis of the ellipse fit to each particle image.
From the particle number concentration (average particle count per unit volume), the volume fraction $\phi_V$ is estimated approximating each particle as a sphere of diameter $d_s$.

\begin{table}
	\centering
	\begin{tabular}{c c c c c}
		Dataset & Mean diameter $d_s$ & Aspect ratio $s_2/s_1$ & Number & $\phi_V \times 10^{-7}$ \\
		& (mm) & (-) & concentration & (-) \\
		& & & (m$^{-3}$) & \\
		& & & & \\
		Jan2016 & $1.09\pm 0.45$ & $0.73 \pm 0.11$ & 816 & 7.4\\
		Nov2018 & $0.65 \pm 0.41$ & $0.65 \pm 0.16$ & 1644 & 6.3\\
		Jan2019 & $0.39 \pm 0.23$ & $0.57 \pm 0.17$ & 56620 & 44\\
	\end{tabular}
	\caption{Snow particle properties (mean and standard deviation) as measured by DIH for all three datasets.}
	\label{tab:tab3}
\end{table}

\begin{figure}
	\centering
	\includegraphics{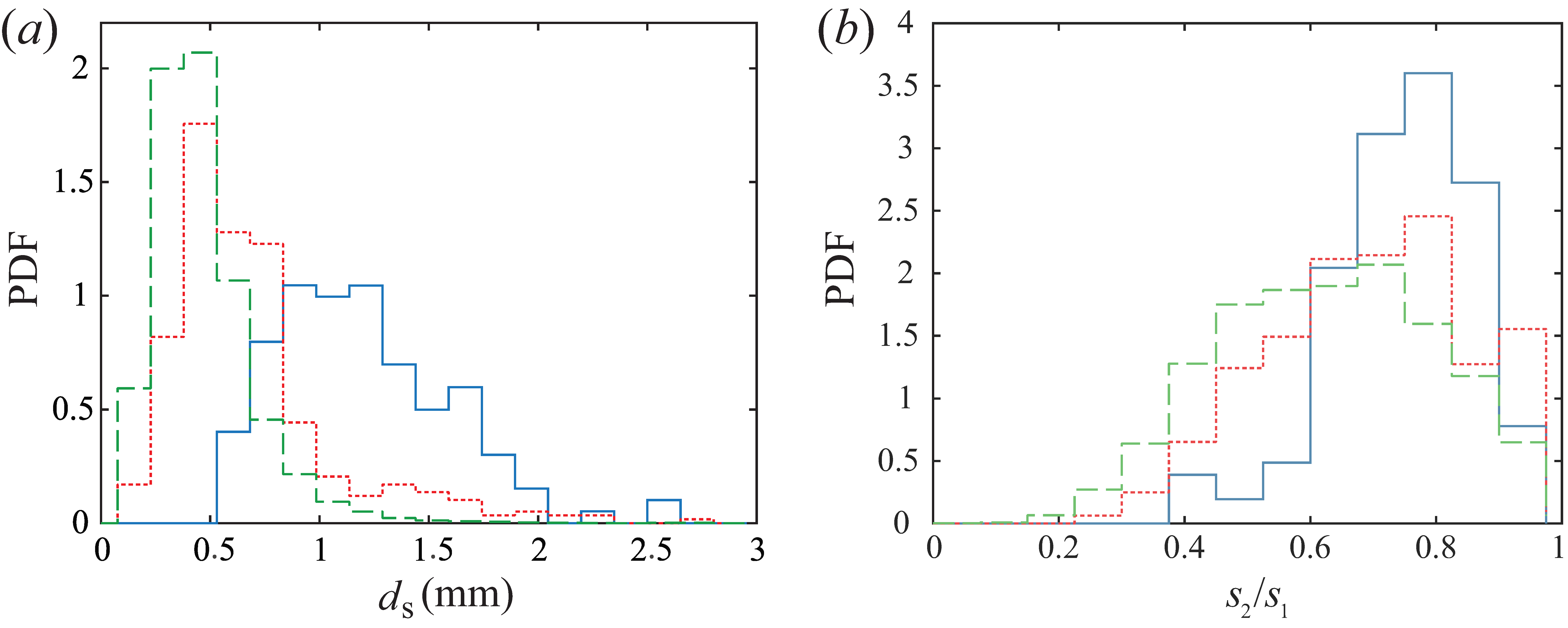}
	\caption{PDFs of (a) size and (b) aspect ratio of the snow particles for Jan2016 (solid blue line), Nov2018 (dotted red line), and Jan2019 (dashed green line). (c) Samples of DIH realizations showing typical hydrometeors.}
	\label{fig:fig5}
\end{figure}

The particle sizes decrease significantly from Jan2016 to Nov2018 to Jan2019, as illustrated by the probability density functions (PDFs) of $d_s$ (figure~\ref{fig:fig5}a).
The distributions of the aspect ratio are similar for the three deployments and indicate relatively compact objects (figure~\ref{fig:fig5}b).
This is confirmed by visual inspection of the DIH realizations: most detected hydrometeors are ice particles and crystals exhibiting moderate level of aggregation and relatively low shape complexity \citep{Garrett2012}.
Given the limited elongation, the influence of the particle anisotropy on the motion dynamics is expected to be small \citep{Voth2017}.
Consistently with the PIV/PTV images of figure~\ref{fig:fig2}, the DIH data for Jan2016 and Nov2018 yields comparable particle concentration, while Jan2019 presents one order of magnitude higher values.

\begin{figure}
	\centering
	\includegraphics{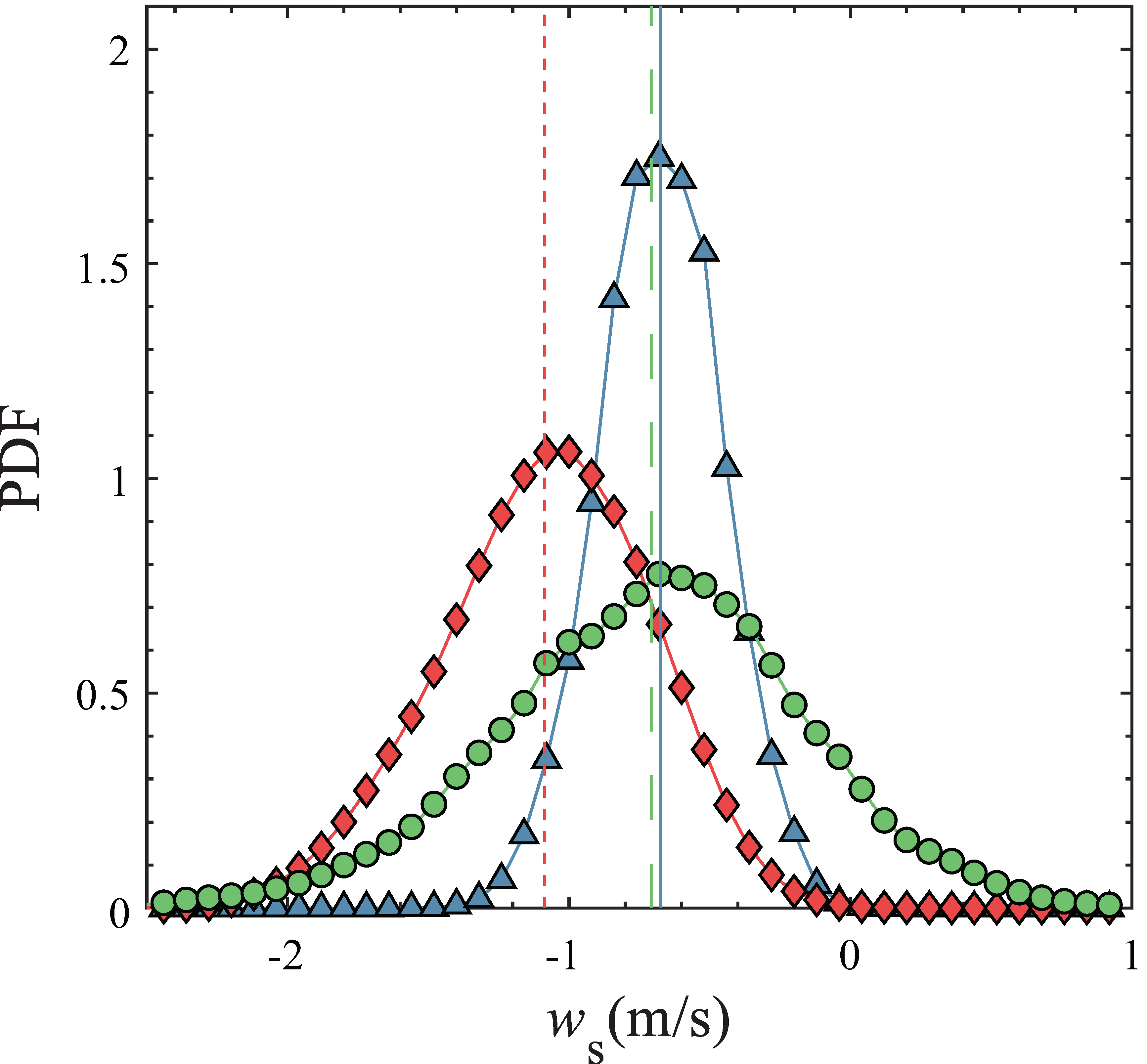}
	\caption{PDFs of the snow particle vertical velocity ($w_s$) as measured by PIV for Jan2016 (blue triangles), Nov2018 (red diamonds), and Jan2019 (green circles). Vertical solid, dotted, and dashed lines mark the mean settling velocity ($W_s$) corresponding to the three datasets, respectively.
	}
	\label{fig:fig6}
\end{figure}

Table~\ref{tab:tab4} reports key statistics for the snow particle vertical velocity $w_s$ and acceleration $a_s$ (where available) obtained from PIV and PTV, respectively.
Angle brackets denote spatio-temporal averaging of the Eulerian fields or ensemble average over all Lagrangian trajectories.
We focus first on the vertical velocities, of which figure~\ref{fig:fig6} shows the PDFs for the three datasets.
The distributions are approximately normal, with only Nov2018 displaying sizeable skewness.
Jan2016 has a similar mean fall speed as Jan2019, but the latter has almost double $w_{s,rms}$, reflecting the increased spread of the velocity distribution, with some particles reaching upward velocities.
Nov2018 shows a mean fall speed almost 60\% higher than the other two cases, and an intermediate $w_{s,rms}$.
A comparison between these velocity distributions and the size distributions in figure~\ref{fig:fig5}a is most interesting.
The trends of both quantities are not reconcilable using classic velocity-diameter relationships; in particular, the fact that snow particles from Nov2018 fall much faster than in Jan2016, while being 40\% smaller in average.
Also, the much larger diameter variance in Jan2016 is at odds with its relatively narrow velocity distribution; vice versa, Jan2019 has the largest spread of velocities while having the narrowest size distribution.

\begin{table}
	\centering
	\begin{tabular}{c c c c c c c}
		Dataset & $W_s = \overline{w_s}$ & $w_{s,rms}$ & $Re_s$ & $|W_s|/u_{rms}$ & $\overline{a_s}$ & $a_{s,rms}$ \\
		& (m/s) & (m/s) & (-) & (-) & (m/s$^2$) & (m/s$^2$)\\
		& & & & & &\\
		Jan2016 & -0.68 & 0.21 & 54.9 & 4.25 & -0.012 & 0.505\\
		Nov2018 & -1.09 & 0.37 & 54.5 & 2.87 & -0.067 & 0.442\\
		Jan2019 & -0.71 & 0.55 & 23.1 & 0.60 & N/A & N/A\\
	\end{tabular}
	\caption{Snow particle properties as measured by DIH for all three datasets.}
	\label{tab:tab4}
\end{table}

To explain these seemingly counterintuitive results, one may speculate that the snow particles in the datasets have significantly different densities, which could account for the incongruence between the distributions of sizes and fall speeds.
However, widely accepted relations between hydrometeor mass $m$ and diameter, $m \sim d_s^{b_m}$ (obtained, e.g., by combined in situ imaging and weighing gauges; \citealt{Tiira2016,VonLerber2017}), do not support this view.
For crystals and small aggregates in the present size range, the exponent $b_m$ is usually equal or larger than 2 \citep{Heymsfield2010,Tiira2016,VonLerber2017}, implying that density (${\sim}\:m/d_s^3$) decreases less than linearly with size \citep{Heymsfield2004}.
Neither can non-linear drag explain the behavior, because the snow particle Reynolds number $Re_s = |W_s|d_s/\nu$ is of the same order for the three cases (and almost the same for Jan2016 and Nov2018).

We instead hypothesize that the present behavior is the result of the influence of air turbulence on snow particle settling.
This is certainly consistent with the increasing variance of vertical velocities from Jan2016 ($Re_\lambda = 938$) to Nov2018 ($Re_\lambda = 3545$) to Jan2019 ($Re_\lambda = 9180$), as more intense turbulence is expected to result in larger spread of the snow particle velocities.
Importantly, the differences in mean fall speed can also be explained by the ability of turbulence to enhance the settling velocity of inertial particles.
Specifically, the larger fall speed in Nov2018 compared to Jan2016 may be the consequence of strong (or stronger) preferential sweeping in the former case.
Likewise, stronger preferential sweeping in Jan2019 than in Jan2016 would explain why the former case shows the same mean fall speed as the latter, despite significant smaller snow particle sizes.
Support to this hypothesis is lent by the particle acceleration data and the correlation between concentration and vertical velocity, presented in the following.

\subsection{Snow particle acceleration}
\begin{figure}
	\centering
	\includegraphics{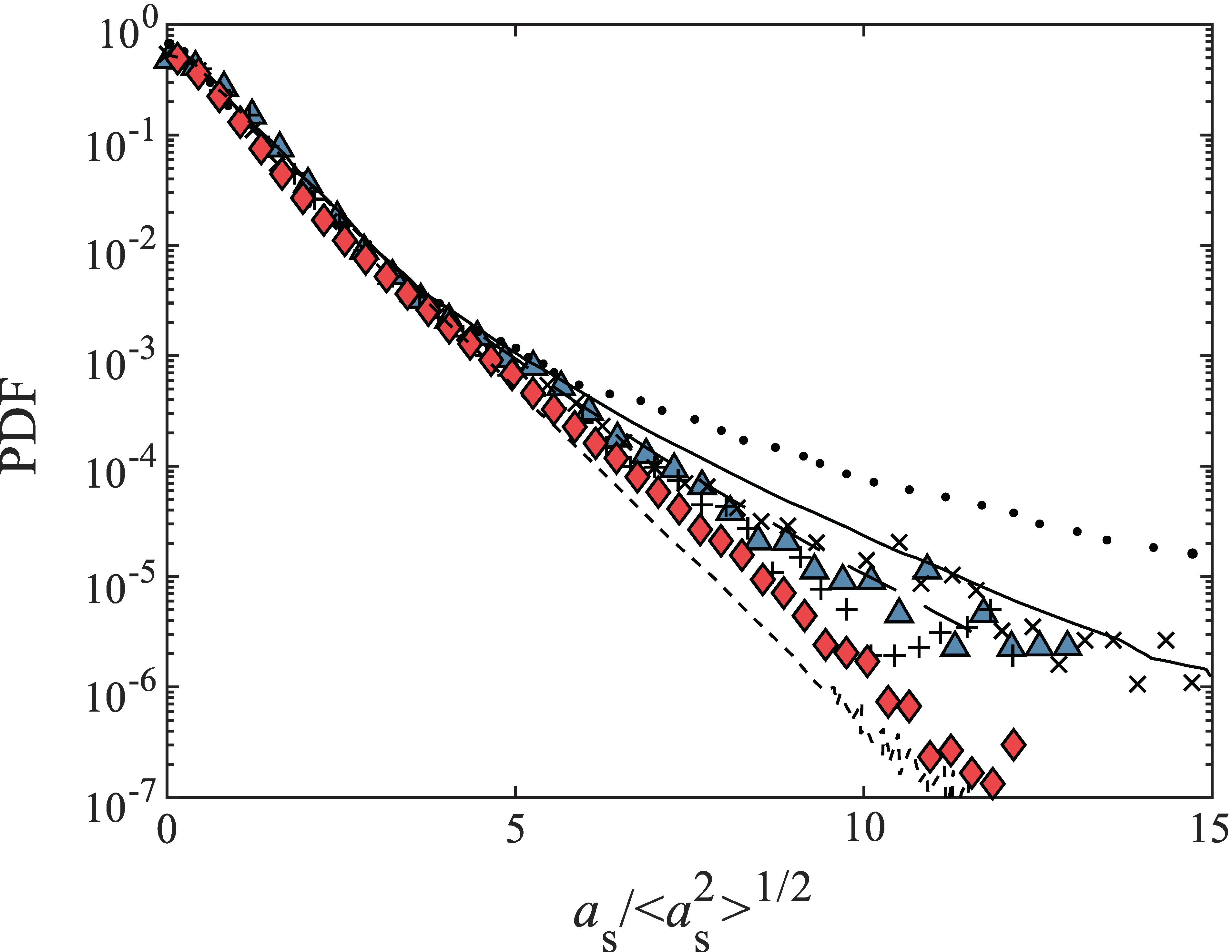}
	\caption{PDFs of horizontal component of snow particle accelerations for Jan2016 (blue triangles) and Nov2018 (red diamonds), compared to $St = 0$ from \citet{Mordant2004} (dots), \citet{Ayyalasomayajula2006} ($St = 0.09$, crosses; $St = 0.15$, plus signs) and \citet{Bec2006} ($St = 0.16$, solid line; $St = 0.37$, dashed line; $St = 2.03$, dotted line).
	}
	\label{fig:fig7}
\end{figure}

Figure~\ref{fig:fig7} shows the PDFs of the fluctuations of the horizontal acceleration for Nov2018 and Jan2016 as obtained by PTV, normalized by their root mean square values.
These are compared to previous numerical and experimental studies of homogeneous turbulence laden with tracers \citep{Mordant2004} and inertial particles of known $St$ \citep{Ayyalasomayajula2006,Bec2006}.
The long exponential tails of the PDFs for all cases highlight the significant intermittency due to intense turbulence events \citep{LaPorta2001,Voth2002}, modulated by the inertia of the particles \citep{Bec2006,Toschi2009}.
As discussed in \citet{Nemes2017}, we can leverage the high sensitivity of the acceleration PDFs to $St$ \citep{Salazar2012} and their low sensitivity to $Re_\lambda$ and the specific flow configuration \citep{Volk2008,Gerashchenko2008}, to estimate the Stokes number of the snow particles.
Without aiming for a precise value, we estimate $St = \mathcal{O}(0.1)$ for Jan2016 and $St = \mathcal{O}(1)$ for Nov2018.
Particles with $St = \mathcal{O}(1)$ are known to display the strongest settling enhancement by preferential sweeping \citep{Wang1993,Aliseda2002,Petersen2019}.
As such, these estimates of $St$ for the snow particles are consistent with the observation of increased fall speeds in Nov2018 compared to Jan2019.

The value of the acceleration r.m.s. (table~\ref{tab:tab4}) provide a further indication that the snow particles captured in Jan2016 are less sensitive to preferential sweeping compared to those in Nov 2018.
In the latter case, the acceleration variance normalized by Kolmogorov scaling is
$a_0 = \overline{a_s'^2}/\left(\varepsilon^{3/2}\nu^{-1/2}\right) = 8.5$.
This value is close to the expectation for tracers at such high $Re_\lambda$ \citep{Ishihara2007,Ireland2016a}, and it results from the action of two opposite effects: particle inertia (which reduces $a_0$ compared to tracers, \citealt{Bec2006}, \citealt{Ireland2016a}), and the effect of particle trajectories crossing the trajectories of fluid elements due to gravitational drift (which increases $a_0$, \citealt{Ireland2016b}, \citealt{Mathai2016}).
The gravitational drift is measured by the ratio of the fall speed to the air velocity fluctuations, $|W_s|/u_{rms}$.
The Jan2016 case has both lower particle inertia (according to our estimates of $St$) and larger gravitational drift through the turbulence (table~\ref{tab:tab4}), and indeed it sees a multifold increase of the non-dimensional variance, $a_0=37.9$.
A similarly dramatic rise of $a_0$  was recently reported for both heavy particles \citep{Ireland2016b} and bubbles \citep{Mathai2016} in homogeneous turbulence, and was explained by the gravitational drift causing the particles to quickly decorrelate from the local turbulence structures, thus experiencing fast-changing fluid motions.
This process limits the ability of the particles to obey preferential sweeping, and therefore this mechanism is expected to be less influential for Jan2016 than Nov2018.

\subsection{Snow particle concentration} \label{sec:sec3.3}
The snow particle concentration fields provide further support to the previous arguments.
Because the Jan2019 dataset does not allow for locating individual particles, we characterized the concentration using the local and instantaneous image intensity $I(x,y,t)$.
This approach is based on the observation that scattered light intensity varies linearly with the particle number density $N$ for monodisperse particles \citep{Bernard2002} and with $N\overline{d_s}^2$ for polydisperse particles \citep{Raffel2018}, and it is often used to measure relative concentration in particle-laden flows (e.g., \citealt{Lai2016}).
We thus calculate the relative concentration as $C^* = I/\overline{I}_{1\mathrm{min}}$, where $\overline{I}_{1\mathrm{min}}$ is the 1-minute moving average of the image intensity at each location.
This normalization helps attenuate temporal fluctuations of the lighting conditions (e.g., due to the power fluctuation of the search light) and does not affect the observed trends.
We also note that using particle counting for Jan2016 and Nov2018 leads to the same conclusions for those datasets.

\begin{figure}
	\centering
	\includegraphics{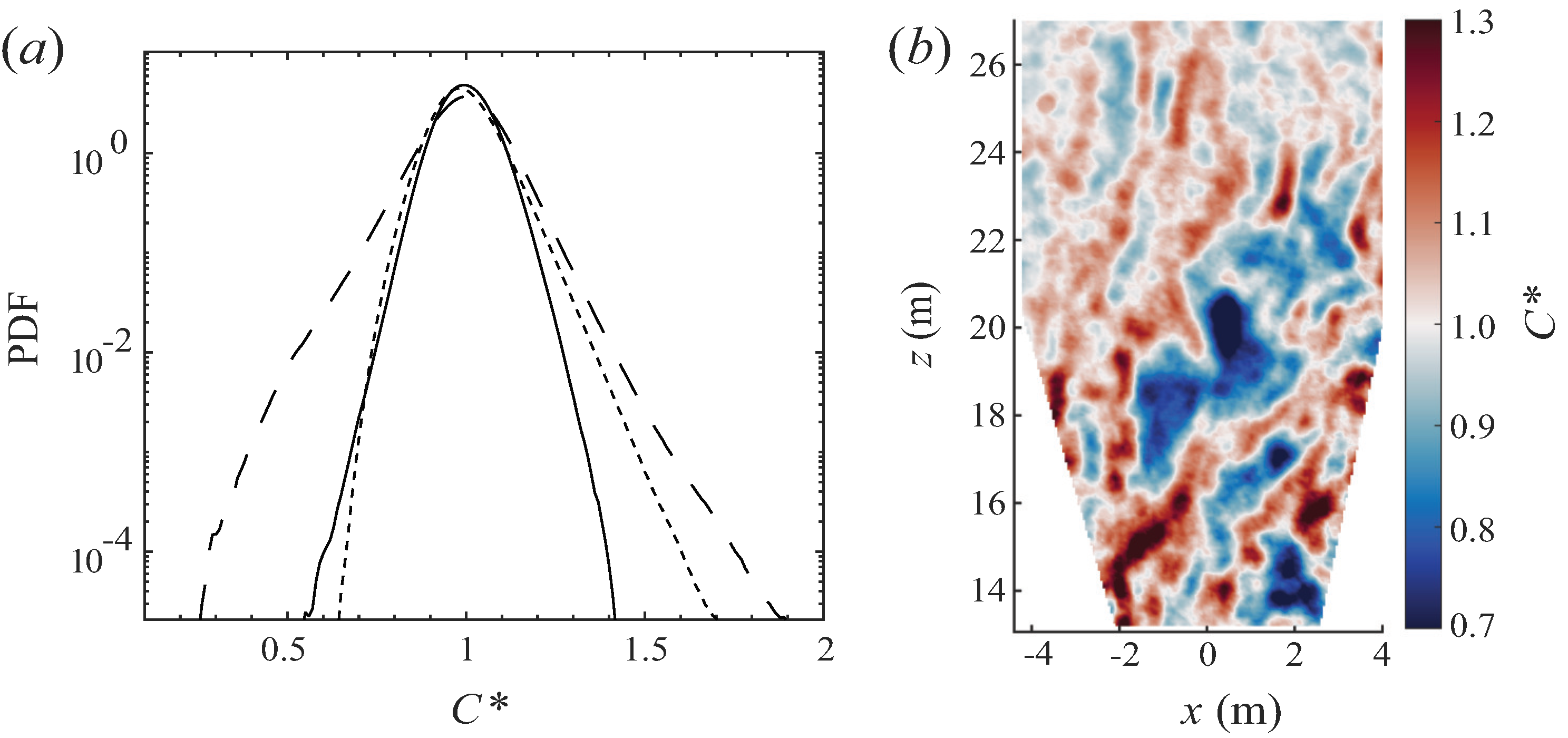}
	\caption{(a) PDF of snow particle relative concentration $C^*$. Black solid, dotted, and dashed lines correspond to the Jan2016 ($Re_\lambda = 938$), Nov2018 ($Re_\lambda = 3545$), and Jan2019 ($Re_\lambda = 9180$), respectively. (b) Instantaneous $C^*$ field from Jan2019.
	}
	\label{fig:fig8}
\end{figure}

Figure~\ref{fig:fig8}a presents the PDF of $C^*$ for the three datasets, indicating different levels of spatio-temporal variability.
Jan2016 displays an approximately Gaussian distribution (with a kurtosis of 3.5, closely matched to the kurtosis of 3 from a Gaussian distribution); while Jan2019 exhibits stretched exponential tails (kurtosis of 4.9), pointing to the significant likelihood of exceptionally low-/high-concentration events.
The trend across the cases parallels that of $Re_\lambda$: the more intense the turbulence, the higher the variance and intermittency in the concentration fields.
Figure~\ref{fig:fig8}b illustrates a sample $C^*$ field from Jan2019, for which the standard deviation of the concentration exceeds 10\% of the mean.
This representative snapshot clearly shows relatively dense zones, vertically elongated and interleaved with more dilute ones.
We will characterize such spatial clustering in the next section.
Here we stress that the more turbulent cases display stronger clustering, and that the latter is typically concurrent with settling enhancement by preferential sweeping \citep{Aliseda2002,Baker2017,Petersen2019,Momenifar2020}.

\begin{figure}
	\centering
	\includegraphics{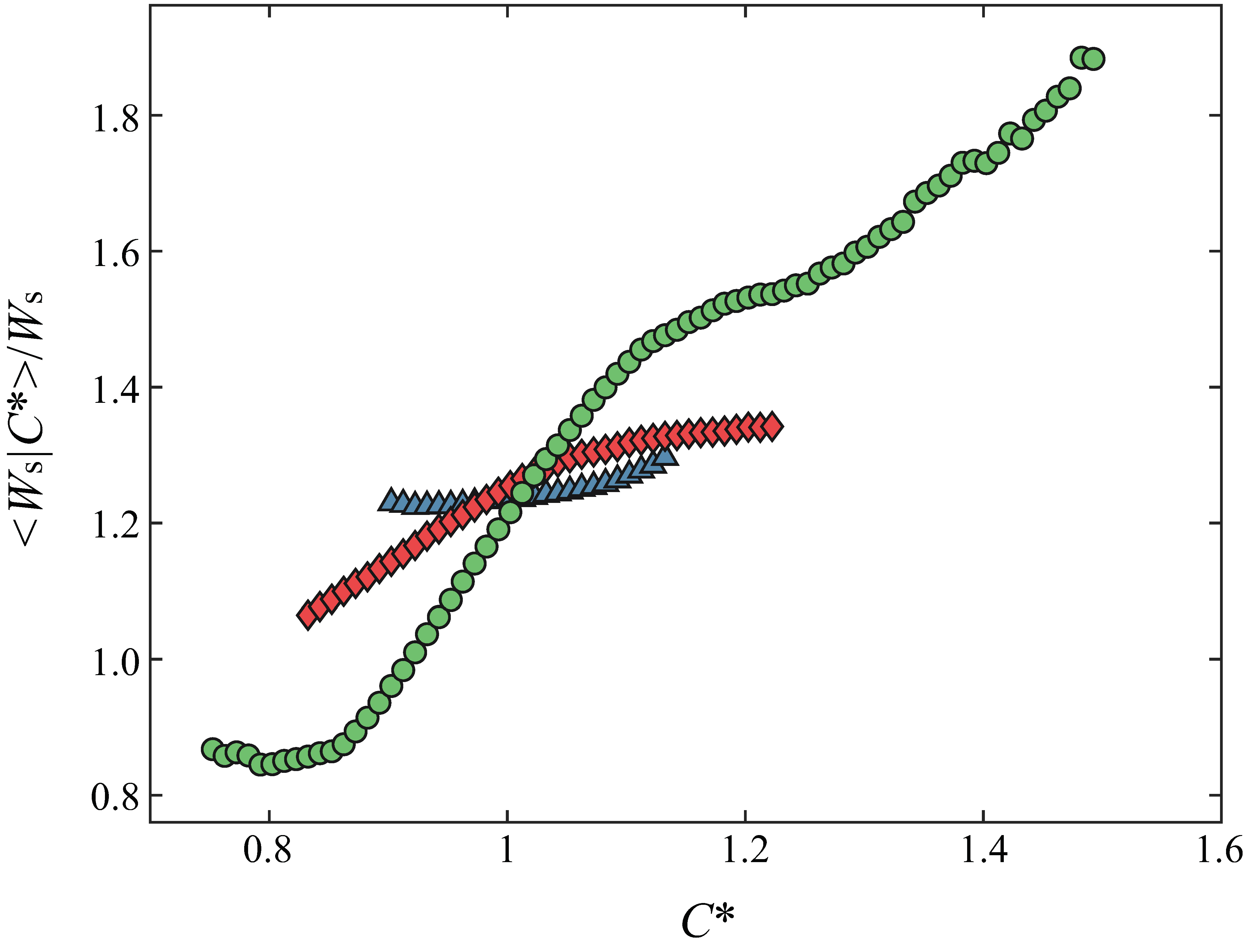}
	\caption{Ensemble-averaged snow particle settling velocity conditioned on value of the local relative concentration $C^*$ and normalized by the unconditional mean settling velocity. Symbols as in figure~\ref{fig:fig6}.
	}
	\label{fig:fig9}
\end{figure}

Direct evidence of preferential sweeping in the more turbulent datasets is provided by the correlation between the local concentration and the simultaneous vertical velocity.
In figure~\ref{fig:fig9} we plot the PIV-based vertical velocity of the snow particles conditioned by the local concentration.
Because the latter is available at every pixel, we use its spatial average in each PIV interrogation window.
The level of correlation between concentration and fall speed is marginal for Jan2016, significant for Nov2018, and the strongest for Jan2019.
Analogous trends were reported since the first demonstration of preferential sweeping by \citet{Wang1993}, and recently confirmed in various simulations and laboratory experiments (among others, \citealt{Aliseda2002,Baker2017,Huck2018,Petersen2019}).
This is a strong indication that the relatively large settling velocity of Nov2018 and Jan2019 (compared to diameter-based expectations) is due to the snow particles preferentially sampling downward air flow regions.

\subsection{Snow particle clustering}
In the above we have shown evidence that snow particles respond to air velocity fluctuations similarly to small inertial particles in turbulence.
It is then of interest to characterize the appearance of one of the most striking effects of particle-turbulence interaction: spatial clustering.
In the following we describe the properties of clusters identified using analogous approaches as in previous laboratory studies.
While quantitative comparisons are thwarted by the uncertainty on the snow particles Stokes number, we show that the concentration fields display the hallmark features repeatedly reported in laboratory studies.

\begin{figure}
	\centering
	\includegraphics{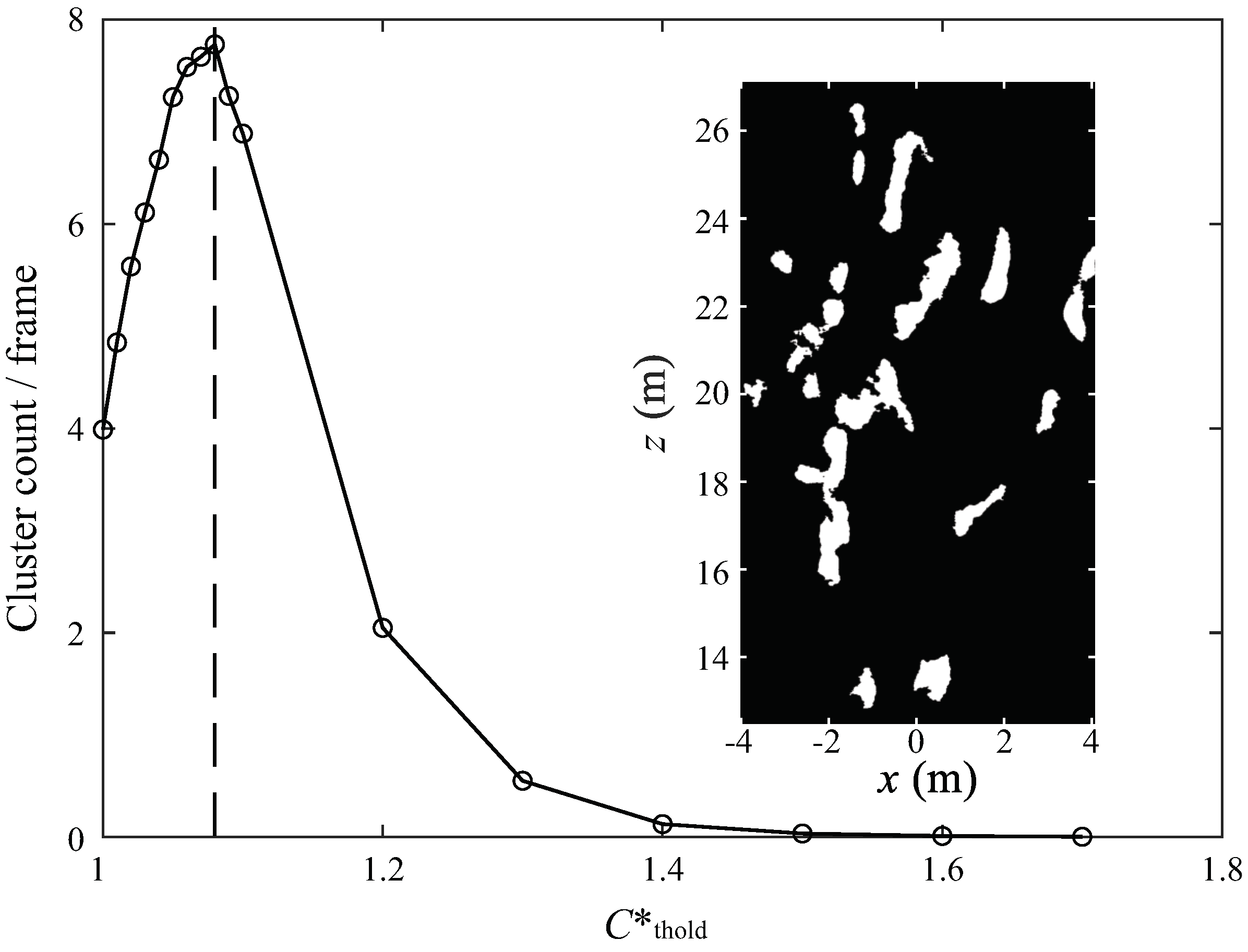}
	\caption{Average number of clusters per image as a function of relative concentration threshold $C^*_\mathrm{thold}$. The inset shows clusters in a binarized concentration field using the threshold that maximizes the number of detected clusters (vertical dashed line in the plot).
	}
	\label{fig:fig10}
\end{figure}

We focus specifically on the Jan2019 case, which presents the most intense turbulence and the most inhomogeneous concentration.
We follow \citet{Aliseda2002} and identify clusters as connected regions where the concentration $C^*$ is above a prescribed threshold.
This approach is standard in image object segmentation, and it has been applied to passive scalars, enstrophy, and velocity fluctuations to detect coherent flow structures in turbulence \citep{Catrakis1996,Moisy2004,Lozano-Duran2012,Carter2018}.
In order to select an objective threshold $C_\mathrm{thold}^*$, we analyse the percolation behaviour of the identified objects \citep{Moisy2004}.
For higher values of the threshold only a few small clusters are detected, which grow in size and number up to a maximum as the threshold is lowered.
They then start to merge, their number decreasing until a single macro-cluster occupies the entire domain.
This process is illustrated in figure~\ref{fig:fig10}, highlighting the chosen threshold which corresponds to the maximum number of identified clusters \citep{Lozano-Duran2012,Carter2018}.
We disregard those that touch the image border, as their full spatial extent could be underestimated.

\begin{figure}
	\centering
	\includegraphics{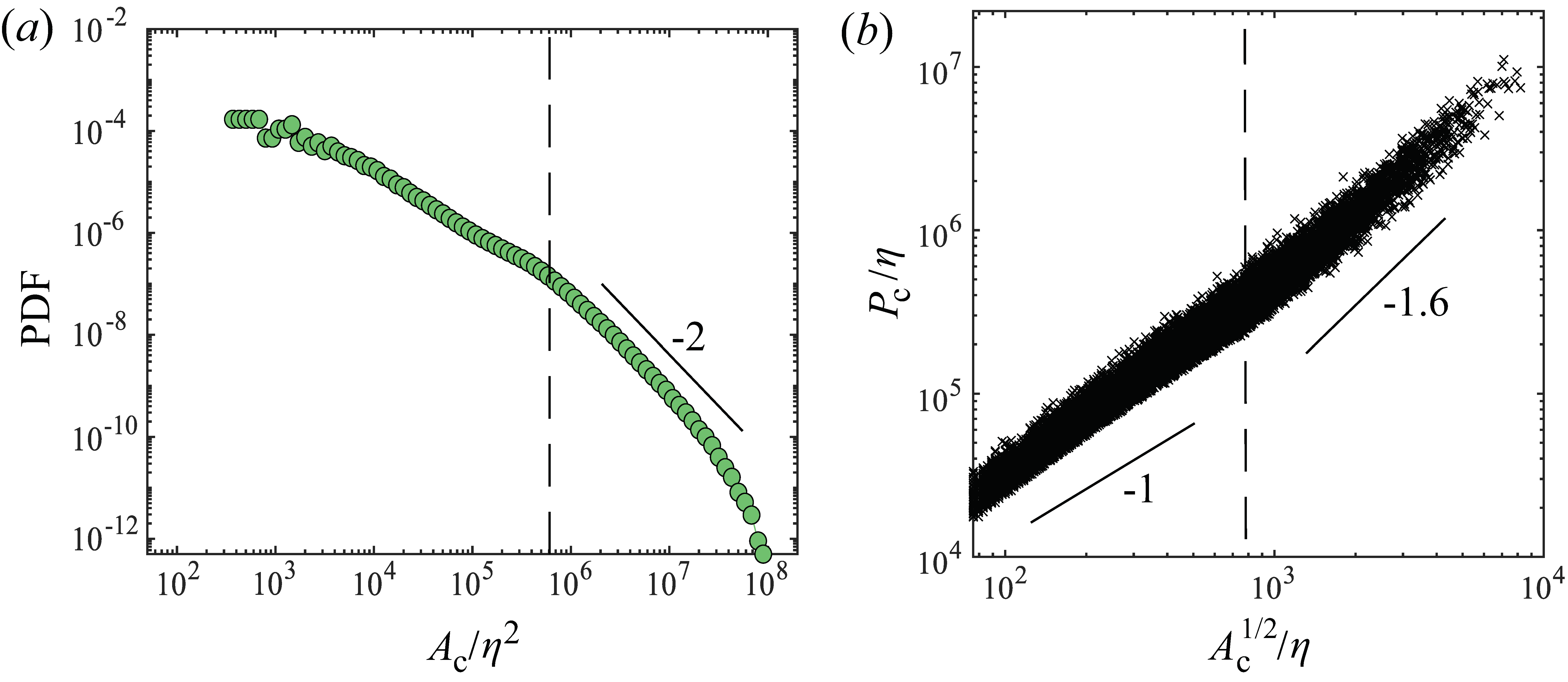}
	\caption{(a) PDF of snow particle cluster area normalized by Kolmogorov scaling, showing a power-law decay with an exponent close to -2 for sizes larger than the light sheet thickness (vertical dashed line). (b) Scatter plot of cluster perimeter versus square root of the cluster area, both normalized by Kolmogorov scaling. 
	}
	\label{fig:fig11}
\end{figure}

Similar to previous imaging studies of particle-laden turbulence, we consider the PDF of the cluster area $A_c$ normalized by the Kolmogorov scale (figure~\ref{fig:fig11}a).
The slope displays a marked change above a length scale corresponding to the light sheet thickness (below which the likelihood of imaging overlapping objects prevents their characterization).
Larger clusters show a power-law decay of the area distribution over more than a decade.
This suggests self-similarity between clusters of different sizes, pointing to their origin from turbulent eddies (which are also self-similar, \citealt{Moisy2004}; \citealt{Baker2017}).
The power-law exponent is consistent with the previously reported value of -2 \citep{Monchaux2010,Petersen2019}.
The detected objects can reach linear dimensions of several meters.
While this challenges the classic estimates of inertial particle clusters being $\mathcal{O}(10\eta)$ in size \citep{Eaton1994}, there is mounting experimental evidence that the cluster size increases with the flow Reynolds number \citep{Sumbekova2017}.

To further describe the cluster topology, figure~\ref{fig:fig11}b shows a scatter plot of their perimeters versus the square root of their areas.
For small clusters, the data points follow a power law with unit exponent, as expected for regular two-dimensional objects.
Such a trend is inherently impacted by the light sheet thickness.
For larger ones (especially for sizes far larger than the light sheet thickness) the exponent is significantly higher, indicating a convoluted structure of the cluster borders.
This trend was observed in several previous imaging studies of particle-laden turbulence \citep{Aliseda2002,Monchaux2010,Petersen2019} and is consistent with the view of inertial particle clusters as fractal sets \citep{Calzavarini2008}.

\begin{figure}
	\centering
	\includegraphics{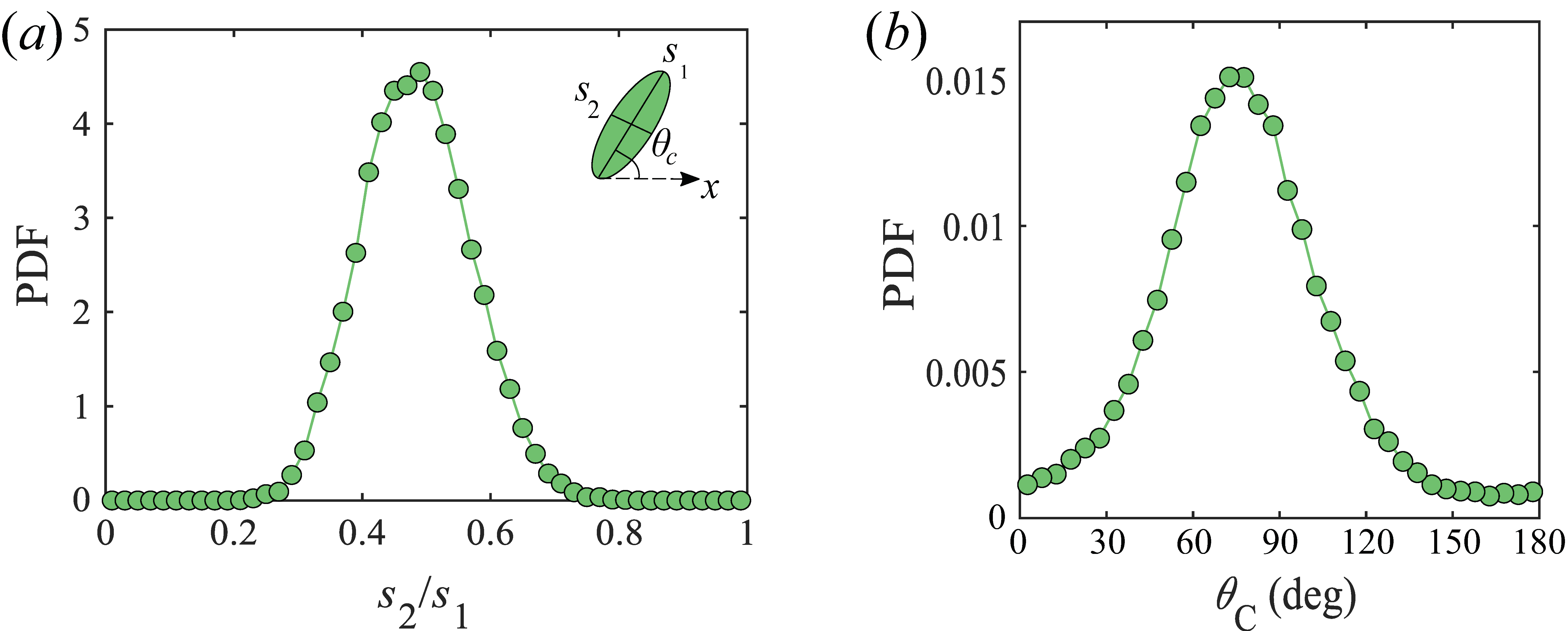}
	\caption{PDFs of (a) aspect ratio and (b) orientation angle of snow particle clusters.
	}
	\label{fig:fig12}
\end{figure}

Clusters of heavy particles settling in turbulence are known to be elongated and preferentially aligned with the vertical direction \citep{Woittiez2009,Dejoan2013,Ireland2016b,Baker2017,Petersen2019}.
Figure~\ref{fig:fig12}a reports the PDF of the cluster aspect ratio (obtained by ellipse-fitting as for the snow particle images from DIH).
The peak slightly below 0.5 is consistent with the laboratory study of \citet{Petersen2019}, who found the distribution to be robust for a range of physical parameters.
The PDF of the angle $\theta_c$ made by the ellipse major axis with the horizontal (figure~\ref{fig:fig12}b) confirms a strong prevalence of vertically oriented clusters.
The weak prevalence of less than 90 degree clusters is likely the consequence of the significant wind speed and mean shear.

\begin{figure}
	\centering
	\includegraphics{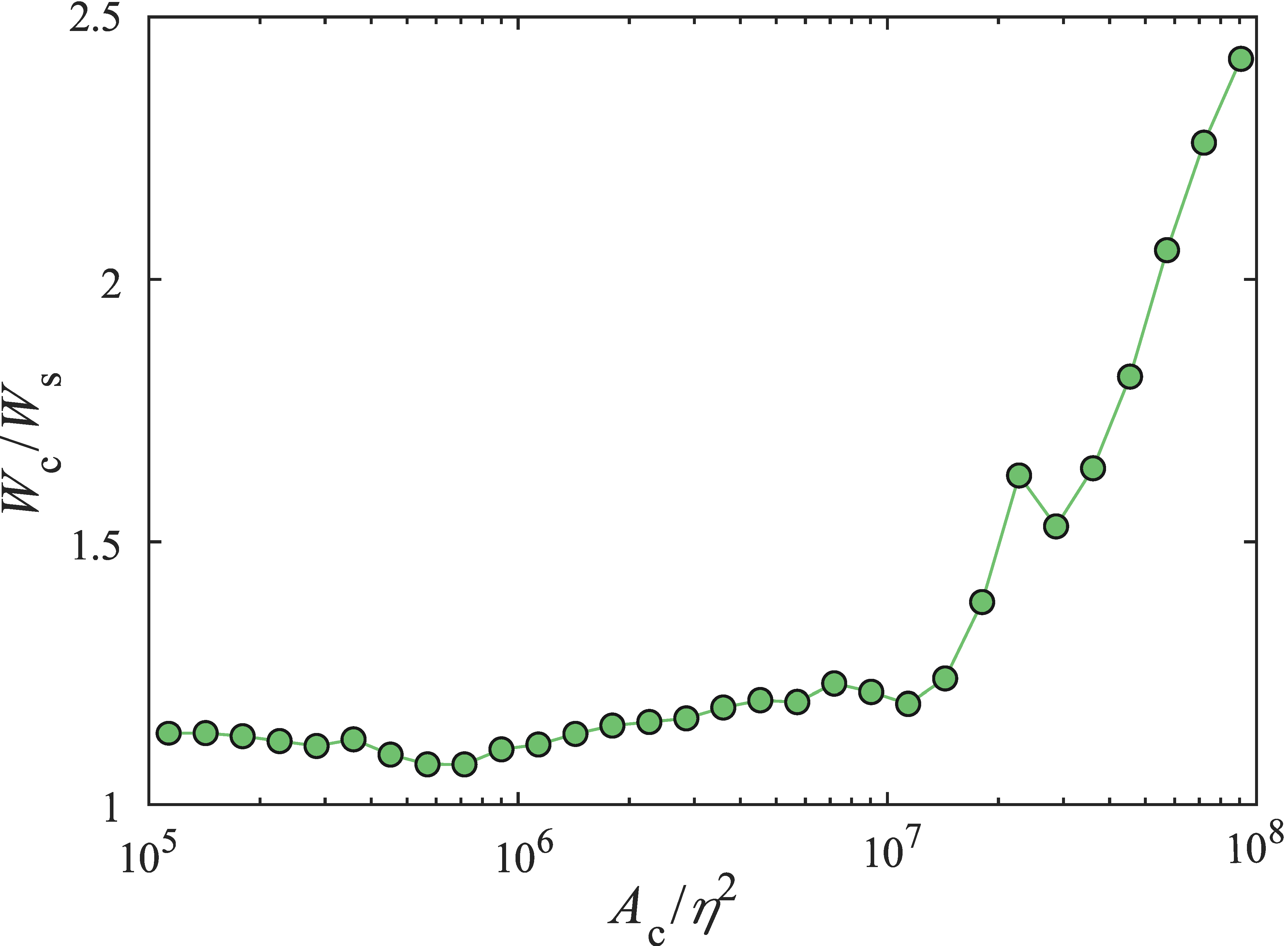}
	\caption{Normalized cluster velocity as a function of cluster size.
	}
	\label{fig:fig13}
\end{figure}

Finally, we consider the cluster fall speed $W_c$, obtained by averaging the vertical velocity at all locations belonging to a given cluster.
This is then ensemble-averaged over all clusters of a certain size, normalized by the mean settling velocity $W_s$, and plotted against the cluster area (figure~\ref{fig:fig13}).
Overall, clusters fall significantly faster than $W_s$, in keeping with the velocity-concentration correlation shown above.
Also, there is an apparent trend of increasing fall speed with cluster size, as seen in laboratory experiments \citep{Huck2018,Petersen2019}.
While this could be merely due to the preferential sampling of downward flow regions, the sharp increase of $W_c$ above a certain cluster size suggests that a more complex interaction between the snow and the turbulent air may be taking place.

The estimated particle volume fraction for Jan2019 is above $10^{-6}$ (table~\ref{tab:tab3}) and it may approach $10^{-5}$ in the denser clusters.
According to widespread criteria for gas-solid flows \citep{Elghobashi1994,Balachandar2010}, in this range of concentrations the dispersed phase is expected to exert a significant back-reaction on the carrier fluid (so-called two-way coupling).
Therefore, the bigger clusters, by virtue of the simultaneous action of large numbers of particles, may be able to collectively drag air down with them, enhancing their fall speed beyond what granted by one-way-coupling mechanisms like preferential sweeping.
Such an effect has been proposed to explain settling velocities observed at similar volume fractions in experiments \citep{Aliseda2002,Huck2018} and numerical simulations \citep{Bosse2006,Frankel2016}.
In fact, the s-shape of the velocity versus concentration curve in figure~\ref{fig:fig9} resembles the recent results of \citet{Huck2018}, who explained their findings using collective particle effects. 

\section{Conclusions} \label{sec:sec4}
The present results provide evidence that atmospheric turbulence affects not only the variance, but also the mean of the snow settling velocity.
Specifically, we show that seemingly contradictory trends between snow particle diameters and fall speeds can be explained by the ability of turbulence to enhance settling.
This effect is attributed to the preferential sweeping of particles into downward regions of the air flow, which is known to be most intense for particles with Stokes number $St = \mathcal{O}(1)$ based on Kolmogorov scaling.
The effect depends therefore on the coupling between the hydrometeor properties and the atmospheric turbulence.
This explanation is consistent with the observed acceleration distributions, from which we infer the range of $St$ for different datasets.
We record large acceleration variance for the case with a fall speed substantially larger than the air velocity fluctuations.
We deduce that, for these snow particles, the crossing trajectory effect (caused by gravitational drift) dominates over the preferential sampling effect (due to the particle inertia): these hydrometeors quickly drift away from the local turbulent structures and are not strongly clustered by them.
On the other hand, preferential sweeping is strong when turbulence fluctuations are comparable to gravitational drift.
The clearest demonstration of preferential sweeping is found in the vertical velocity conditioned by the local concentration: regions of high concentration display higher settling velocities.
This might also imply a back-reaction of the particles on the flow through collective drag.
Our estimates of the highest volume fractions are indeed above classic thresholds for two-way coupling, although one should exert caution in applying those to a flow laden with complex particles.
We have also demonstrated that, in the cases where strong preferential sweeping is inferred, the concentration field is highly non-uniform.
Clusters appear over a wide range of scales, displaying signature features identified in laboratory experiments and numerical simulations: power-law size distribution, fractal-like silhouette, vertical elongation, and large fall speed that increases with size.

Taken together, these results confirm and extend the conclusion of \citet{Nemes2017}: the phenomenology of inertial particles in turbulence, built over decades of canonical flow studies, is largely applicable to the dynamics of snow settling in air.
Presently, none of these well-known concepts are incorporated in weather forecasting models.
Other recent studies, also imaging-based, demonstrated that those concepts are in fact directly applicable to atmospheric flows: within clouds, clustering of droplets was recently observed using airborne holographic instruments \citep{Beals2015,Larsen2018}, and elongated regions devoid of droplets were identified at a mountain-top station \citep{Karpinska2019}.

Environmental flows give access to much larger ranges of scales than laboratory or numerical experiments, enabling the exploration of fundamental fluid mechanics questions.
To our knowledge, the present field work represent the highest-Reynolds number flow measurements in which inertial particle clustering is observed and quantified.
The fact that we observe very large clusters (up to the integral scales of the turbulence) lends support to recent claims that these clusters grow larger with $Re_\lambda$ \citep{Sumbekova2017}, and that increasing $Re_\lambda$ extends the range of scales to which the particles respond \citep{Tom2019}.
Of course, the non-canonical aspects of naturally occurring particle-laden flows have to be considered.
In particular, the morphology of snow particles is expected to play an important role, especially for complex-shaped and elongated snow particles \citep{Westbrook2017}.
Our understanding of the interaction of anisotropic particles with turbulence has seen tremendous progress in recent years \citep{Voth2017}, and imaging studies capable of testing these dynamics in the field are warranted.
This will require three-dimensional imaging at high spatial and temporal resolution, and will benefit from novel capabilities now available for Lagrangian tracking (see e.g., \citealt{Guala2008a} and the recent review by \citealt{Discetti2018}).

An important aspect which the present study cannot directly address is represented by the hydrometeor collision rate, and the impact turbulence has on it.
This is expected to be strongly related to the polydispersity of the snow particles.
Polydispersity drives the classic (gravitational) mechanism by which larger particles fall faster than and collide with smaller ones \citep{Pruppacher1997}.
Recent simulations, however, show that turbulence enhances the relative velocity of polydisperse particles also in the horizontal direction \citep{Dhariwal2018}.
Fundamental studies in this area are needed to extend the applicability of particle-turbulence dynamics to environmental flows.
Additionally, field studies are warranted to investigate the correlation between the level of collision-driven aggregation and atmospheric turbulence.

\section*{Acknowledgements}
The authors gratefully acknowledge the support of the US National Science Foundation through grant NSF-AGS-1822192. We also thank engineers from St. Anthony Falls Laboratory, including James Tucker, James Mullin, Chris Ellis, Jeff Marr, Chris Milliren and Dick Christopher, for their assistance in the experiments.

\section*{Declaration of Interests}
The authors report no conflict of interest.

\bibliographystyle{jfm}

\end{document}